\documentclass[preprintnumbers,showpacs,amsmath,amssymb,floatfix,9pt,prd,twocolumn,
superscriptaddress,nofootinbib]{revtex4}
\usepackage{latexsym}
\usepackage{epsfig}
\usepackage{amssymb}
\begin{document}

\title{Charged compact star model in Einstein-Maxwell-Gauss-Bonnet gravity}

\author{Piyali Bhar}
\email{piyalibhar90@gmail.com
 } \affiliation{Department of
Mathematics,Government General Degree College, Singur, Hooghly 712 409, West Bengal,
India}

\author{Megan Govender}
\email{megandhreng@dut.ac.za}
\affiliation{Department of Mathematics, Faculty of Applied Sciences, Durban University of Technology, Durban, South Africa
}

\begin{abstract}
Present paper provides a model of static charged anisotropic fluid sphere
in the Einstein–Maxwell–Gauss–Bonnet (EMGB) theory of gravitation. We select KB {\it anstz} as the metric co-efficient along with electric field intensity. To develop our model we assume a linear EoS between the radial pressure and the matter density. The model obtained here is found to
satisfy the elementary physical requirements like non-negativity nature of pressure and density, the energy conditions, continuity of the metric co-efficients,  subliminal velocity of sound and also pressure free hypersurface etc. It may also noted that the solution is free from any kind of singularities.

\end{abstract}

\maketitle

\section{Introduction}

Einstein's classical general relativity (CGR) has been instrumental in probing the nature of the gravitational field and its associated dynamics. From the standard Big Bang model of the universe through to inhomogeneous cosmological models, Quasi-steady state cosmologies, phantom cosmologies, emergent Universes and dark energy models, CGR has sucessfully accounted for various observable phenomena such as the origin of the Cosmic Microwave Background radiation (CMBR), the acceleration of the Universe, Hubble rate, nucleosynthesis amongst others. On the astrophysics frontier, CGR has been successfully used to model compact objects such as neutron stars, white dwarfs, puslars and black holes. Theorists have been richly rewarded with a plethora of physics and insights into the nature of matter at ultra-high densities. The Cosmic Censorship Conjecture has been thoroughly probed under various assumptions of the initial state of the collapsing matter, spacetime symmetry and the presence of dissipation. Cosmic censorship in higher dimensions has produced interesting results as compared to their 4-d counterparts.

Although highly successful and more robust than Newtonian gravity, Einstein's classical general relativity has been shown to display various shortfalls, particularly in the quantum regime. CGR breaks down in the vicinity of the initial Big Bang singularity. The cosmological coincidence problem has plagued cosmologists employing CGR. Likewise in astrophysics, the information paradox problem, behavior of spacetime in the vicinity of a singularity, to highlight just a few shortcomings warrants the need for an extension of Einsteinian gravity.

The Randall-Sundrum braneworld scenario has attracted widespread attention amongst astrophysicists and cosmologists for the past decade or more. In the braneworld construction gravity is the only fundamental interaction which propagates through the bulk and this affects the dynamics on the brane. Contributions from the bulk via projection of the Weyl stresses lead to modification of the Einstein field equations. In a cosmological scenario brane effects lead to the modified Friedmann equations, the solutions of which compensate for shortcomings in classical 4-D Einstein gravity. Modeling stars on the brane have yielded interesting results in terms of compactness, stability and the role played by dissipation. The exterior solution for a static star on the brane is not unique as in the 4-D case. It is well-known that the collapse of a radiating star in classical GR requires that the pressure on the boundary of the star is nonvanishing. The exterior is no longer empty and is described by the Vaidya solution. In the Braneworld collapse model, the exterior of a collapsing sphere is mediated by a radiation zone which in turn matches to the Schwarzschild solution. The radiation zone is necessary because of the non-zero Weyl stresses across the hypersurface which divides the interior and exterior spacetimes. The Motola-Mazur gravastar model has been investigated within the framework of Braneworld gravity. The requirement that the pressure is anisotropic within the core of the gravastar occurs naturally in the Braneworld scenario. The anisotropy is a result of the Weyl tidal effects.

Recently, there has been a surge in the interest of modified theories of gravity which have as a lower limit 4-D classical general relativity. The propagation of gravity in higher dimensions have yielded rich insights into the kinematical and dynamical properties of the gravitational interaction. The Einstein-Gauss-Bonnet (EGB) gravity, in particular, has generated lots of interest amongst researchers. The modeling of compact objects in EGB gravity has shown that compactness, redshift and stability are modified when compared to their 4-D counterparts.  The Buchdahl inequality in EGB gravity softer than its Einsteinian limit. Cosmic censorhip has been investigated in EGB gravity revealing interesting properties of continued gravitational collapse. The universality of the Schwarzschild constant density sphere has been demonstrated in EGB gravity.

In a recent paper Bhar et al. have carried out a comparative study of compact objects in classical 5-D gravity and EGB gravity. The interior matter configuration consisted of a neutral perfect fluid with anisotropic stresses ($p_r \neq p_t$).  They showed that contributions from the Gauss-Bonnet terms lead to modifications in the compactness, stability and redshift of compact objects. It is well-known that the electromagnetic field naturally arises in classical 5-D Einstein gravity. The kinematics and dynamics of the electromagnetic field were studied by Kaluzua and Klein. In this paper we study  the role played by the electromagnetic field on compact objects within EGB gravity. We seek to ascertain the effect of charge on the properties of charged compact objects within classical 5-D Einstein gravity versus 5-D EGB gravity theory.

This paper is stuctured as follows: In section II we outline the framework of EGB gravity. The modified Einstein-Maxwell field equations with a nonzero Gauss-Bonnet term are presented in section III. Using the ansatz of Bhar et al. we generate a model of an anisotropic charged star within EGB gravity in section IV. The exterior spacetime and matching conditions are presented in section V. In section VI we present a detailed physical analysis of our model.

\section{Einstein-Maxwell-Gauss-Bonnet Gravity}

It is well-known that the Gauss-Bonnet action in five dimensions takes the form
\begin{equation}\label{1}
S=\int \sqrt{-g} \left[\frac{1}{2}(R +\alpha
L_{GB})\right]d^5x +S_{matter},
\end{equation}
where the Gauss-Bonnet coupling constant, $\alpha \geq 0$ gives a measure of the string tension in arising in string theory. A pleasing feature of the action (\ref{1}) is that while the
Lagrangian is quadratic in the Ricci tensor, Ricci scalar and the Riemann tensor, the resulting equations of motion are second
order quasi-linear. Gauss-Bonnet effects arise in higher dimensional models for $n > 4$.\par
In presence of electric field, the EMGB field equations can be cast as
\begin{equation}\label{2}
G_{ab}+\alpha H_{ab}=\Upsilon_{ab}=T_{ab}+E_{ab},
\end{equation}
where $G_{ab}$ represents the Einstein
tensor.
where $\Upsilon_{ab}$ is the total energy-momentum tensor, $T_{ab}$ and $E_{ab}$ are respectively the energy–momentum tensor corresponding to
matter and the electromagnetic field,
The Lanczos tensor is given by
\begin{eqnarray}\label{3}
H_{ab}&= &2\left(R R_{ab}-2R_{ac}R^c_b- 2
R^{cd}R_{acbd}+R^{cde}_aR_{bcde}\right)\nonumber\\
&&-\frac{1}{2}g_{ab}L_{GB},
\end{eqnarray}
where the Lovelock term has the form

\begin{equation}\label{4}
L_{GB}=R^2 +R_{abcd}R^{abcd}- 4R_{cd}R^{cd}.
\end{equation}
In the above formalism we use geometric units with the coupling constant $\kappa$ set to unity.

\section{Field equations}
\label{sec:3}
In order to model the interior spacetime of the compact object we assume a spherically symmetric geometry described by five dimensional line element
\begin{eqnarray}
\label{5} ds^{2}& =& -e^{2\nu(r)} dt^{2} + e^{2\lambda(r)} dr^{2} +
 r^{2}(d\theta^{2} + \sin^{2}{\theta} d\phi^2\nonumber\\
 && +\sin^{2}{\theta} \sin^{2}{\phi^2}
 d\psi^2),
\end{eqnarray}
where the metric functions $e^{\lambda}$ and $e^{\nu}$ encode the bevaviour of the gravitational field.

The energy-momentum tensor of fluid the distribution and electromagnetic
field are defined respectively as 
\begin{equation}
{T^i}_j = [(\rho + p_t)v^iv_j - p_t{\delta^i}_j + (p_r - p_t) \theta^i \theta_j],\label{matter}
\end{equation}

\begin{equation}
{E^i}_j = \frac{1}{4}(-F^{im}F_{jm} + \frac{1}{4}{\delta^i}_jF^{mn}F_{mn}).\label{electric}
\end{equation}
where $v^i$ is the four-velocity, $v^i=e^{\nu(r)/2}{\delta^i}_4$, $\theta^i$ is a unit space-like
vector in the radial direction, $\theta^i =
e^{\lambda(r)/2}{\delta^i}_1$, $\rho $ is the energy density,
$p_r$ is the radial pressure and and $p_t$ is the tangential pressure. The components for ${T^i}_j$ and ${E^i}_j$
are defined respectively as:
\begin{equation}
 {T^1}_1=-p_r,\, {T^2}_2={T^3}_3=-p_t,\, {T^4}_4=\rho
 \end{equation}
 \begin{equation}
 {E^1}_1=-{E^2}_2=-{E^3}_3={E^4}_4=\frac{1}{8\,\pi}\,e^{\nu+\lambda}\,F^{14}\,F^{41}.\\
\end{equation}

On the grounds of spherical symmetry, the
four-current component is only a function of radial distance, $r$.
The only non vanishing components of electromagnetic field tensor
are $F^{41}$ and $F^{14}$, related by $F^{41} = - F^{14}$, which
describes the radial component of the electric field.
If $q(r)$ represents the total charge contained within the
sphere of radius $r$, then it can be defined by the relativistic Gauss law as
\begin{equation}
q(r) = 4\pi \int_0^r \sigma r^2 e^{\lambda/2} dr = r^2 \sqrt{-F_{14}F^{14}}. \label{charge}
\end{equation}
From Eq.(\ref{charge}), we obtain
\begin{equation}
F^{41}=- e^{-(\nu+\lambda)/2}\,\frac{q(r)}{r^2}.
\end{equation}
 The Einstein-Maxwell-Gauss-Bonnet (EMGB) field equations yield
\begin{eqnarray} \label{emgb}
\label{7a}\kappa T_0^0=\kappa\rho+E^2&=& \frac{3}{e^{4\lambda }r^3} \left(4\alpha
\lambda'
+re^{2\lambda}-re^{4\lambda}\right.\nonumber\\
&&\left.- r^2 e^{2\lambda}\lambda' -4\alpha e^{2\lambda}\lambda'\right),\\
\label{7b}-\kappa T_1^1=\kappa p_r-E^2& = &\frac{3}{e^{4\lambda }r^3}
\left[
(r^2 \nu' +r +4\alpha \nu')e^{2\lambda}\right.\nonumber\\
&& \left.-re^{4\lambda}-4\alpha \nu'\right] , \\
\label{7c} -\kappa T_2^2=\kappa p_t+E^2&=& \frac{1}{e^{4\lambda }r^2} \left[- e^{4\lambda
}- 4\alpha \nu''+ 12 \alpha \nu' \lambda' \right.\nonumber\\
&&\left.-4 \alpha
(\nu')^2\right] +\frac{1}{e^{2\lambda }r^2} \left[1- r^2 \nu' \lambda' \right.\nonumber\\
&&\left.+2r \nu'-2r  \lambda' +r^2(\nu')^2 +r^2 \nu'' \right.\nonumber\\
&&\left.-4\alpha
\nu'\lambda' + 4\alpha (\nu')^2 +4\alpha \nu''\right]\nonumber\\
&&=-\kappa T_3^3.
\end{eqnarray}
where $\kappa=8\pi$ and $(\prime)$ denotes differentiation with respect to the radial coordinate $r$.  In writing the EMGB field equations we have taken the gravitational constant $G$ and the velocity of light $c$ to be unity. We will now turn our attention to seeking an exact solution to the highly nonlinear system of equations (\ref{7a})--(\ref{7c}).

\begin{figure}[htbp]
    \centering
        \includegraphics[scale=.8]{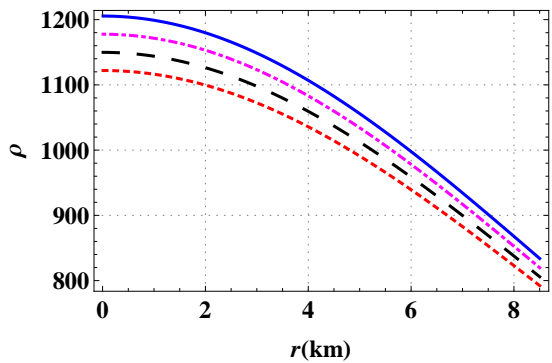}
       \caption{Matter density is plotted against the radial distance r. The description of the curves is as follows: red small dashed line, black medium dashed line, brown long dashed line, pink dot-dashed line and blue solid line for $\alpha = 0$, $\alpha= 10$,a= 15,a= 30 and a = 50, respectively by fixing A = 0.0062 and different values of $B$ mentioned in Table 1}
    \label{rho}
\end{figure}

\begin{figure}[htbp]
    \centering
        \includegraphics[scale=.8]{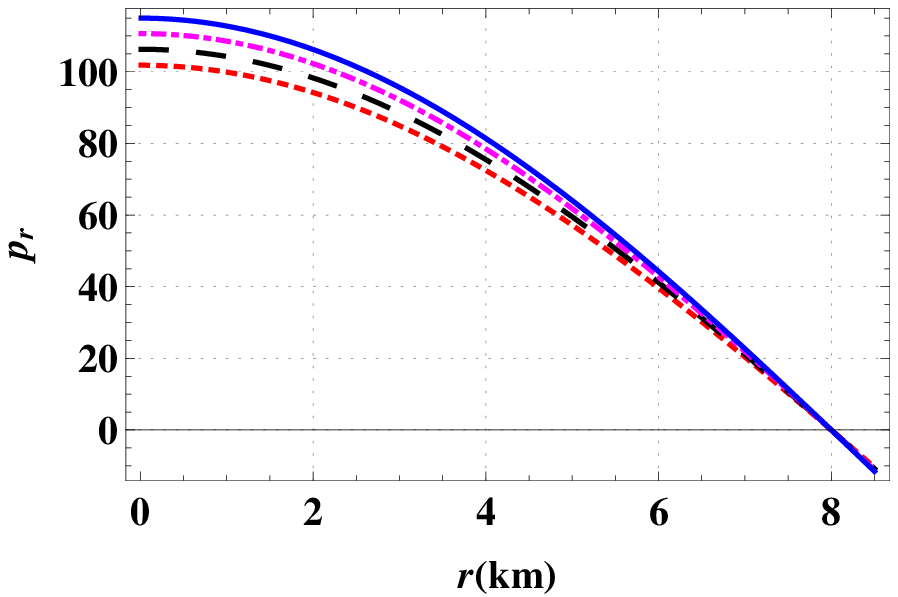}
       \caption{Radial pressure $p_r$ shown against the radial distance r. The description of the curves is the same as in Fig.\ref{rho}
}
    \label{pr}
\end{figure}

\begin{figure}[htbp]
    \centering
        \includegraphics[scale=.8]{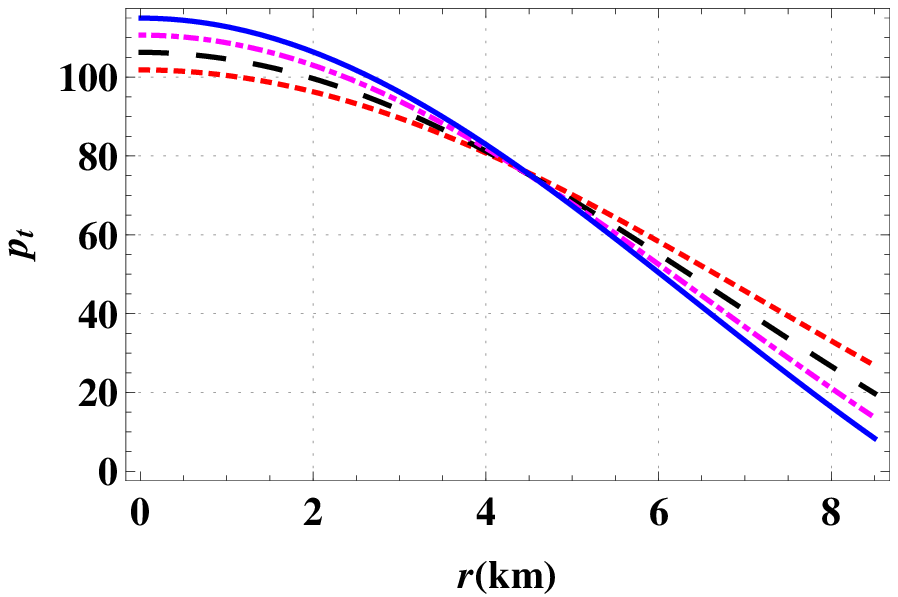}
       \caption{Transverse pressure $p_t$ shown against the radial distance r.The description of the curves is the same as Fig.\ref{rho}}
    \label{pt}
\end{figure}

\section{Anisotropic charged model}
The role of charge and anisotropy in modeling compact objects in classical general relativity has been emphasized in several recent studies. Our intention here is to generate an exact model to the set of equations (\ref{7a})-(\ref{7c}). We observe that the EMGB equations contain five unknowns namely $\rho,\,p_r,\,p_t,\,\lambda,\,\nu$ and $E^2.$ This clearly indicates that we have freedom in choosing either the metric potentials or matter content ab intio. \par Recently, there have been successful attempts at modeling compact objects using the Krori-Barua ansatz \cite{kb} given by
\begin{equation}\label{an}
2\lambda(r)=Ar^{2}~~~~\text{and}~~~~ 2\nu(r)=Br^{2}+C
\end{equation}
where the constants $A,\,B$ have dimensions of $length^{-2}$ and $C$ is a dimensionless constant. These constants are determined from requiring that the interior spacetime match smoothly to the exterior spacetime. We should point out that Junevicus\cite{jun} adopted an interesting approach which determined these
constants in terms of
the physical quantities mass, charge and radius of the
source.\par
KB metric is a nice platform to model the compact star since it provides a singularity free model. Several works have done by the researchers based on the KB {\em ansatz} \cite{pb1,pb2,pb3,pb4,pb5,fr1,fr2,fr3,fr4}.

Using this ansatz given in eqn.(\ref{an}) the field eqns.(\ref{7a})-(\ref{7c}) can be rewritten as,
\begin{eqnarray}
\rho+\frac{E^2}{8\pi} &=& \frac{3e^{-2Ar^{2}}}{8\pi r^{2}}\left[-4A\alpha+e^{2Ar^{2}}\right.\nonumber\\
&&\left.+e^{Ar^{2}}\left(4A\alpha+Ar^{2}-1\right)\right]\label{a1}
\\
p_r -\frac{E^2}{8\pi} &=& \frac{3e^{-2Ar^{2}}}{8\pi r^{2}}\left[-4\alpha B-e^{2Ar^{2}}\right.\nonumber\\
&&\left.+e^{Ar^{2}}\left(1+4\alpha B+Br^{2}\right)\right]\label{a2}
\\
 p_t+\frac{E^2}{8\pi}& =&\frac{e^{-2Ar^{2}}}{8\pi r^{2}}\left[-e^{-2Ar^{2}}-4\alpha B\left\{1+(B-3A)r^{2}\right\}\right]\nonumber\\
&&+\frac{e^{-Ar^{2}}}{8\pi r^{2}}\Big[1+4\alpha B+\left\{B(3+4\alpha B)\right.\nonumber\\
&&\left.-2A(1+2\alpha B) \right\}r^{2}+B(B-A)r^{4}\Big]\label{a3}
\end{eqnarray}
The anisotropic factor $\Delta=p_t-p_r$ assumes the following form

\begin{eqnarray}
\Delta&=&\frac{e^{-2Ar^{2}}}{8\pi r^{2}}\left[2e^{2Ar^{2}}+4\alpha B \left\{2+(3A-B)r^{2}\right\}\right]\nonumber\\&&-\frac{e^{-Ar^{2}}}{8\pi r^{2}}\Big[2+
2Ar^{2}+(A-B)Br^{4}\nonumber\\
&&+4\alpha B \left\{2+(A-B)r^{2}\right\}\Big]-\frac{E^2}{4\pi}
\end{eqnarray}

Along with KB-metric we also assume that the radial pressure $p_r$ is related with the matter density $\rho$ by a linear equation of state
\begin{equation}\label{eos}
p_r=\beta \rho-\gamma
\end{equation}
where $\beta$ and $\gamma$ are positive constants.\\
Adding eqns.(\ref{a1}) and (\ref{a2}) we get,
\begin{equation}\label{1}
\rho+p_r=\frac{3 (A + B) e^{-2 A r^2}}{r^2}\big[-4 \alpha + e^{A r^2} (4 \alpha + r^2)\big]
\end{equation}
Solving eqn.(\ref{1}) with the help of eq.(\ref{eos}), the matter density and radial pressure are obtained as,
\begin{eqnarray}\rho&=&\frac{e^{-2 A r^2}}{8 (1 + \beta) \pi r^2} \left[-12 \alpha (A + B) + 8 \gamma e^{2 A r^2} \pi r^2 \right.\nonumber\\
&&\left.+3 (A + B) e^{A r^2} (4 \alpha + r^2)\right]\label{rho1}\\
   p_r&=&\frac{e^{-2 A r^2}}{8 (1 + \beta) \pi r^2} \left[-12 \alpha (A + B)\beta - 8 \gamma e^{2 A r^2} \pi r^2 \right.\nonumber\\
&&\left.+3 \beta (A + B) e^{A r^2}(4 \alpha + r^2)\right]\label{pr1}
\end{eqnarray}
Plugging the expression of $\rho$ in eqn.(\ref{7a}) the expression for the electric field intensity is obtained as,
\begin{eqnarray}
E^2&=&\frac{e^{-2 A r^2}}{(1 + \beta) r^2} \bigg[12 \alpha (B - A \beta) + e^{2 A r^2} (3 + 3 \beta - 8 \gamma \pi r^2)\nonumber\\&& -
 3 e^{A r^2} \Big\{1 + \beta + (B - A \beta) (4 \alpha + r^2)\Big\}\bigg]
 \end{eqnarray}
and the charge density $\sigma$ can be written explicitly as,
\begin{equation}\label{sig}
\sigma(r)=
\end{equation}

 Consequently the expression for the transverse pressure $p_t$ is obtained as,
 \begin{eqnarray}\label{pt1}
   p_t&=&\frac{e^{-2Ar^2}}{8 (1 + \beta) \pi r^2}\bigg[4 \alpha \Big\{3 A \beta - B (4 + \beta) \nonumber\\
   &&- B (-3 A + B) (1 + \beta) r^2\Big\} -
 4 e^{2 A r^2} (1 + \beta - 2 \gamma \pi r^2) \nonumber\\&&+
 e^{A r^2} \Big\{4 (1 + \beta) + \Big(3 B (2 + \beta) - A (2 + 5 \beta)\Big) r^2 \nonumber\\&&+
    B (-A + B) (1 + \beta) r^4 \nonumber\\&&+
4 \alpha \Big(-3 A \beta + B \big(4 + \beta + (-A + B) (1 + \beta) r^2\big)\Big)\Big\}\bigg]\nonumber\\
\end{eqnarray}

The gravitational and thermodynamical behaviour of our model is completely specified. In order to fix the arbitrary constants arising from the KB ansatz we will match the interior spacetime given by metric (\ref{5}) and the exterior spacetime.

\section{exterior spacetime and boundary condition}
In order to generate a complete model of a bounded configuration we require the smooth matching of the interior spacetime ${\cal M}^{-}$ to the extrior spacetime ${\cal M}^{+}$ across the boundary of the star defined by $r = r_0$ where $r_0$ is a constant. Since the star is not radiating energy the exterior spacetime is empty and is given by the charged analogue of the Boulware-Deser spacetime given by
\begin{equation} \label{exterior}
ds^2 = -{\cal F}dt^2 + \frac{dr^2}{\cal F} + r^{2}(d\theta^{2} + \sin^{2}{\theta} d\phi^2
 +\sin^{2}{\theta} \sin^{2}{\phi^2}
 d\psi^2)
\end{equation}
where ${\cal {F}}(r) = K + \frac{r^2}{4\alpha}\left(1 - \sqrt{1 + \frac{8\alpha M}{r^4} - \frac{8\alpha Q^2}{3r^6}}\right)$ and $K$ is an arbitrary constant. The quantities $M$ and $Q$ represent the gravitational mass and total charge of the fluid sphere as determined by an observer located at infinity.

\section{Physical Analysis of the present charged model}
\begin{itemize}
\item {\bf Regularity of metric co-efficient}\par
For our model the metric potentials take  finite value and singularity free at the center (r = 0), since at the center, $e^{\nu} = C$ and $e^{\lambda}= 1$.
\item {\bf Regularity of electric charge}\par
At the center of the star the electric field is obtained as,
\begin{eqnarray}
E^2(0)=\frac{-3 B + 12 A^2 \alpha \beta + A (3 - 12 \alpha B + 6 \beta) -
 8 \gamma \pi}{1 + \beta},\nonumber\\
 \end{eqnarray}
 Now let us impose the condition $E^2(r=0)=0$ which gives the expression of $\gamma$ as,
\begin{equation}\label{gam}
\gamma = \frac{-3 B + 12 A^2 \alpha \beta + A (3 - 12 \alpha B + 6 \beta)}{8
    \pi}.
\end{equation}
\item{\bf {Regularity of pressure and density}}\par
We have shown the behavior of the matter density in fig.~\ref{rho}. From the figure it is clear that the matter density $\rho$ is monotonic decreasing function of `r' and it takes the positive value inside the star. It is also noted that the central density increases when $\alpha$ takes the higher value.
At the center of the model of the charged star the charge and the matter density are obtained as,
\begin{eqnarray}
  \rho(r=0) &=& \frac{3 A + 12 A^2 \alpha + 3 B + 12 A \alpha B + 8 \gamma \pi}{
8 \pi (1+ \beta)}\label{r}
\end{eqnarray}
Plugging the expression of $\gamma$ of eqn. (\ref{gam}) into (\ref{r}) the central density is obtained as,
\begin{eqnarray}\rho_c&=&\frac{3 A (1 + 2 A \alpha)}{4\pi}
\end{eqnarray}
and consequently the expression for central pressure is obtained as,
\begin{eqnarray}
p_c&=&\alpha \frac{3 A (1 + 2 A \alpha)}{4\pi}-\beta
\end{eqnarray}
The radial and transverse pressure are plotted in fig.~\ref{pr} and fig.~\ref{pt} respectively. The central radial pressure shows the same behavior as it occurs in the the central density, i.e., it takes the higher value as $\alpha$ increases. In case of transverse pressure the surface value admits lower value as $\alpha$ increases and it is clear from the plot of $p_t$ as well as from table~1. In other words we can also say that though the value of transverse pressure differ at the center of the star for different values of $\alpha$ but the the value of transverse pressure coincide at a point irrespective of the value of $\alpha$ (please refer fig.~\ref{pt}).\par
Now by using the condition $p_r(r=R)=0$, the expression for $B$ is obtained as,
\begin{eqnarray}
B=\frac{A \Big[4 \alpha \beta - (4 \alpha + R^2) \beta e^{A R^2}  + e^{2 A R^2} R^2(1+
     2 \beta+
     4 A \alpha \beta)\Big]}{-4 \alpha \beta +e^{A R^2}(
   4 \alpha \beta + \beta R^2) + e^{2 A R^2}R^2(1 +
   4 A \alpha )}.\nonumber\\
   \end{eqnarray}
   Differentiating eqns. (\ref{rho1}),\,(\ref{pr1}) and (\ref{pt1}), the expression for density and pressure gradient are obtained as,
\begin{eqnarray}
\frac{d\rho}{dr}&=&\frac{3 (A + B) e^{-2 A r^2}}{4 (1 + \beta) \pi r^3} \Big[4 \alpha (1 + 2 A r^2)\nonumber\\
&& -e^{A r^2} \Big\{A r^4 + 4 \alpha (1 + A r^2)\Big\}\Big]\label{r1}\\
\frac{dp_r}{dr}&=&\frac{3 \beta(A + B) e^{-2 A r^2}}{4 (1 + \beta) \pi r^3} \Big[4 \alpha (1 + 2 A r^2)\nonumber\\
   &&-e^{A r^2} \Big\{A r^4 + 4 \alpha (1 + A r^2)\Big\}\Big]\label{r2}
\end{eqnarray}
\begin{eqnarray}\label{r3}
\frac{dp_t}{dr}&=&\frac{e^{-2 A r^2}}{4 (1 + \beta) \pi r^3}\bigg[4 (1 + \beta) e^{2 A r^2} +
 4 \alpha \Big\{-3 A \beta + B (4 + \beta) \nonumber\\&&+
    2 A \Big(-3 A \beta + B (4 + \beta)\Big) r^2 +
    2 A B (-3 A + B) \times \nonumber\\&&(1 + \beta) r^4\Big\} +
 e^{A r^2} \Big\{-4 -4 \beta -4 \alpha \big(-3 A \beta + B (4 + \beta)\big) \nonumber\\&&+
    4 A \Big(-1 - \beta -
       \alpha \big(-3 A \beta + B (4 + \beta)\big)\Big) r^2 \nonumber\\&&+ \Big(B^2 (1 + \beta) -
       A B \big(7 + 4 \beta + 4 \alpha B (1 + \beta)\big) \nonumber\\&&+
       A^2 \big(2 + 5 \beta + 4 \alpha B (1 + \beta)\big)\Big) r^4 \nonumber\\&&+
    A (A - B) B (1 + \beta) r^6\Big\}\bigg]
    \end{eqnarray}

\item{{\bf Measurement of the anisotropy}}\\
 The anisotropic factor for our present model is calculated from the following formula,
 \begin{eqnarray}
 \Delta=p_t -p_r.
  \end{eqnarray}
The above equation gives the information about the anisotropic behavior of the present model and it remains positive if $p_t > p_r$. It suggests that the anisotropy being drawn outward and anisotropic force is repulsive in nature. On the other hand, if $p_t < p_r$, the anisotropy turns negative which corresponds to attractive nature of the anisotropic force. In case of our present model, the variations of the anisotropic factor with respect to the radial coordinate r is presented in fig.~\ref{delta} for different values of $\alpha$ and from the figure it is clear that $p_t < p_r$, suggesting that the anisotropic force is repulsive in nature and it is a required condition for the construction of the compact object as proposed by Gokhroo and Mehra \cite{gm}. The anisotropic factor vanishes at the center of the star and $\Delta_{r=R}>0$ \cite{liang}. We can also note that $\Delta_{r=R}$ admits higher value as $\alpha$ decreases.

\begin{figure}[htbp]
    \centering
        \includegraphics[scale=.8]{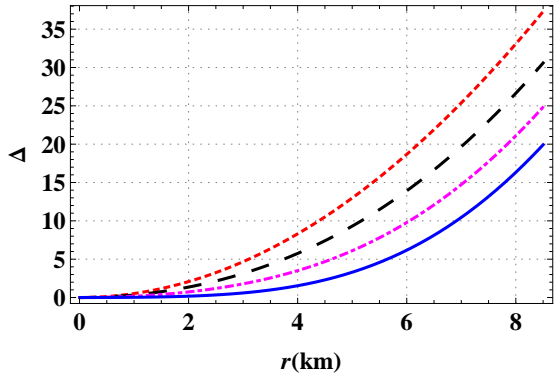}
       \caption{Anisotropic factor is shown against the radial distance r. The description of the curves is the same as Fig.\ref{rho}
}
\label{delta}
\end{figure}

\begin{figure}[htbp]
    \centering
        \includegraphics[scale=.8]{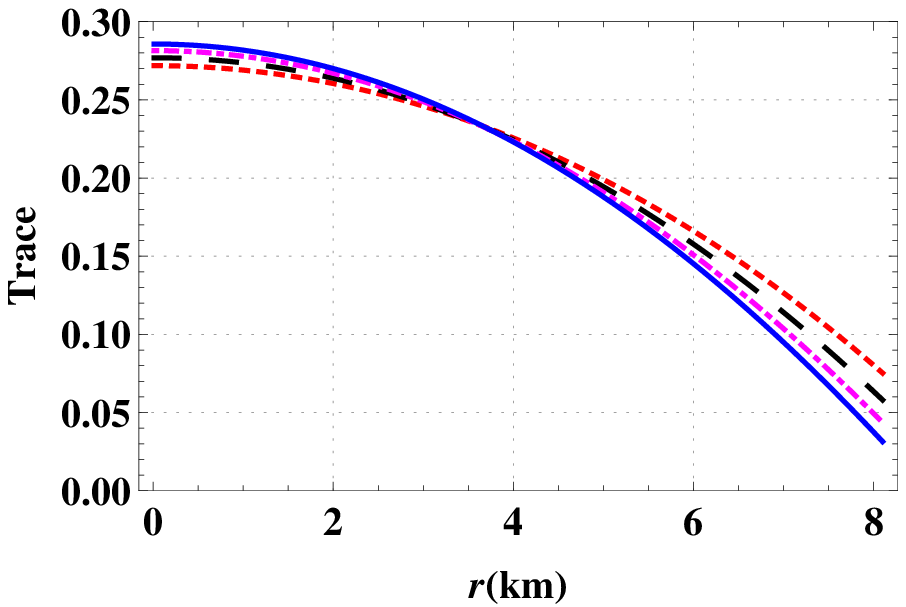}
       \caption{$\frac{p_r+2p_t}{\rho}$ is shown against the radial distance r. The description of the curves is the same as Fig.\ref{rho}}
    \label{fig:12}
\end{figure}

\begin{figure}[htbp]
    \centering
        \includegraphics[scale=.8]{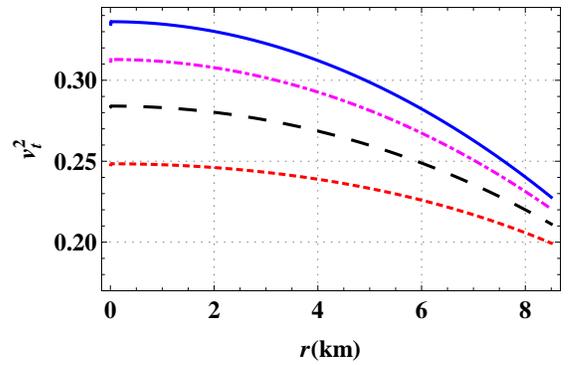}
       \caption{Radial velocity of sound is shown against the radial distance r. The description of the curves is the same as Fig.\ref{rho}}
    \label{vt}
\end{figure}

\begin{figure}[htbp]
    \centering
        \includegraphics[scale=.8]{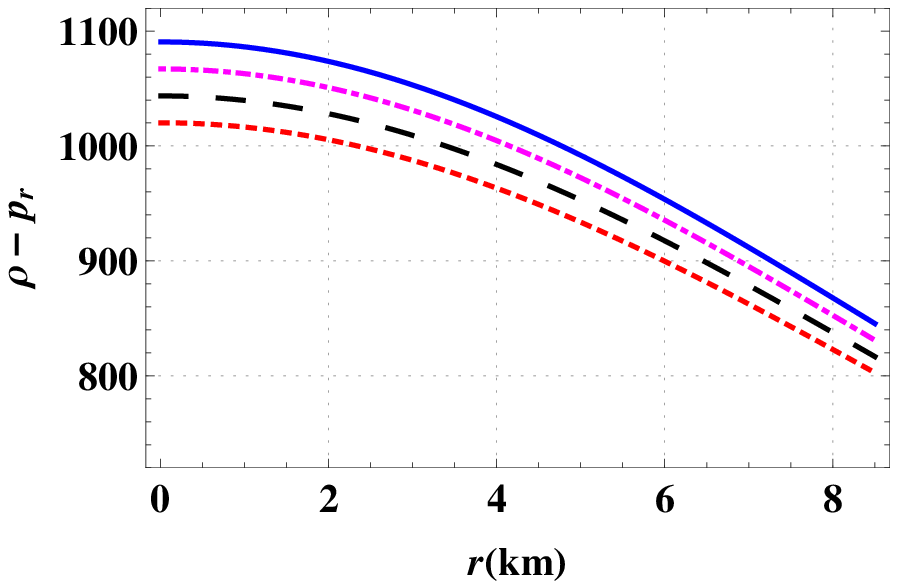}
       \caption{$\rho-p_r$ is shown against the radial distance r. The description of the curves is the same as Fig.\ref{rho}}
    \label{ec1}
\end{figure}

\begin{figure}[htbp]
    \centering
        \includegraphics[scale=.8]{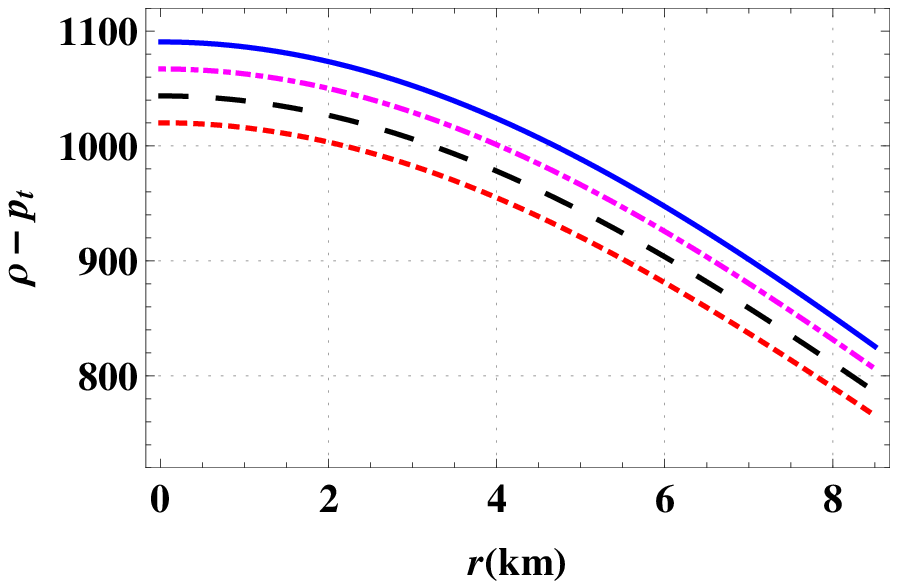}
       \caption{$\rho-p_t$ is shown against the radial distance r. The description of the curves is the same as Fig.\ref{rho}}
    \label{ec2}
\end{figure}

\begin{figure}[htbp]
    \centering
        \includegraphics[scale=.8]{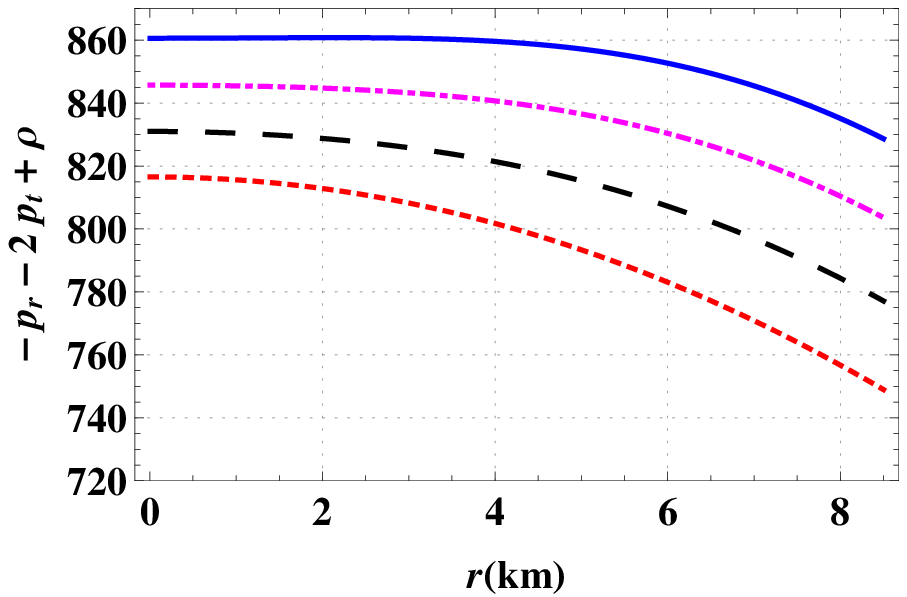}
       \caption{$\rho-p_r-2p_t$ is shown against the radial distance r. The description of the curves is the same as Fig.\ref{rho}}
    \label{ec3}
\end{figure}

\begin{figure}[htbp]
    \centering
        \includegraphics[scale=.8]{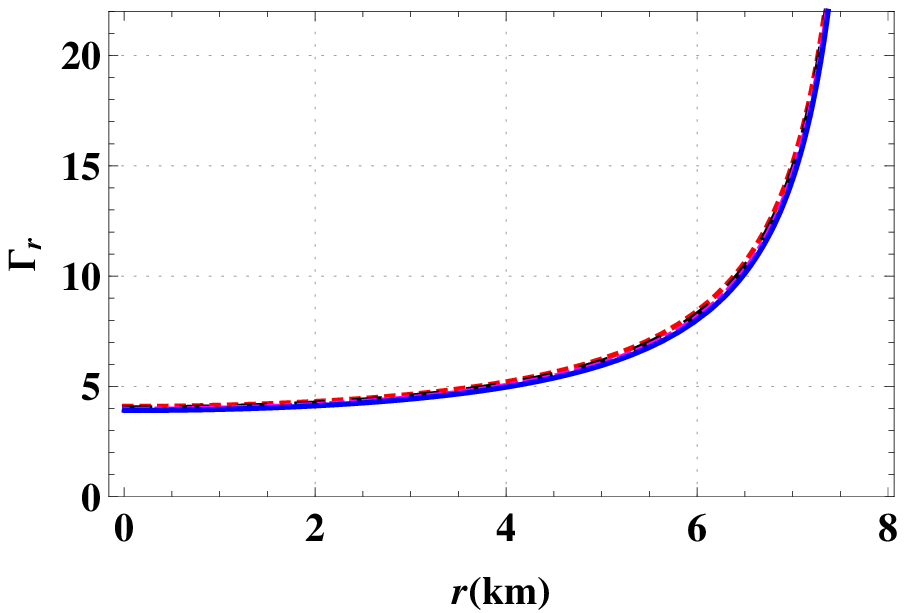}
       \caption{$\Gamma_r$ is shown against the radial distance r. The description of the curves is the same as Fig.\ref{rho}}
    \label{gammar}
\end{figure}

\begin{figure}[htbp]
    \centering
        \includegraphics[scale=.8]{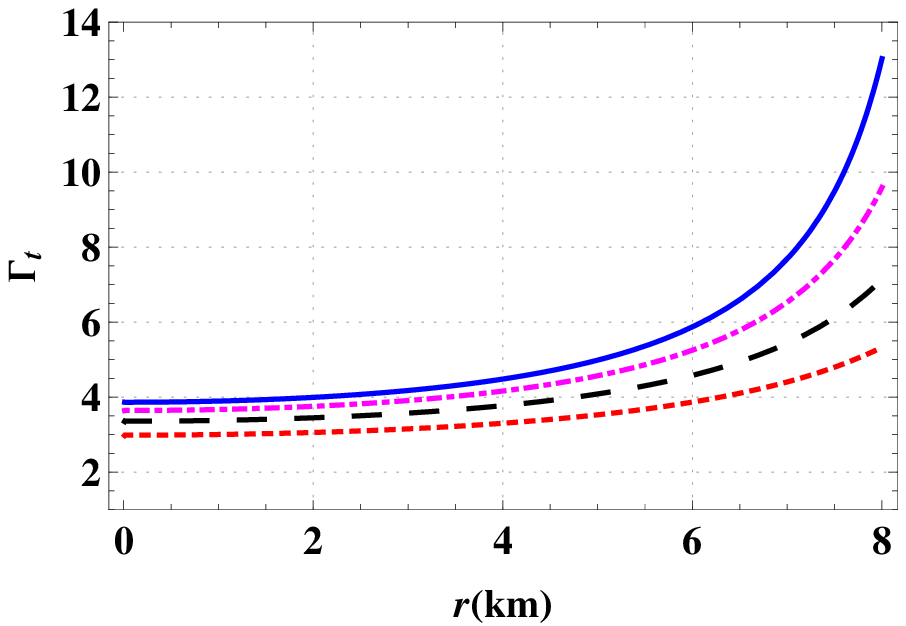}
       \caption{$\Gamma_t$ is shown against the radial distance r. The description of the curves is the same as Fig.\ref{rho}}
    \label{gammat}
\end{figure}

\begin{table*}
\caption{The numerical values of $B$, central density $(\rho_c)$, surface density $(\rho_s)$, central pressure $(p_c)$ and mass $(M_{\odot})$ are obtained by fixing $A=0.0062$ and radius of the star = $8$ km. and different values of the coupling constant $\alpha$}
\label{sphericcase}
\begin{tabular*}{\textwidth}{@{\extracolsep{\fill}}lrrrrrrrl@{}}
\hline
$\alpha$& \multicolumn{1}{c}{$B$} & \multicolumn{1}{c}{$\rho_c$} & \multicolumn{1}{c}{$\rho_s$} & \multicolumn{1}{c}{$p_c$}   \\
& $km^{-2}$ & $gm.cm^{-3}$ & $gm.cm^{-3}$ & $dyne.cm^{-2}$\\
\hline
0& 0.0073239 & $1.99719\times10^{15}$ & $1.46478\times10^{15}$ & $1.62918\times 10^{35}$ \\
2& 0.0070247 & $2.04672\times10^{15}$ & $1.49101\times10^{15}$ & $1.70048 \times 10^{35}$    \\
4& 0.0067516 & $2.09625\times10^{15}$  &$1.51766\times10^{15}$ &$1.77049 \times 10^{35}$ \\
6& 0.0065015 & $2.14578\times10^{15}$ & $1.54467\times10^{15}$ & $1.83939 \times 10^{35}$\\
8& 0.0062715 & $2.19531\times10^{15}$ & $1.57201\times10^{15}$ &$1.90732 \times 10^{35}$\\
10& 0.0060593& $2.24484\times10^{15}$ & $1.59962\times10^{15}$ & $1.97437 \times 10^{35}$\\
12& 0.0058629& $2.29437\times10^{15}$ & $1.62749\times10^{15}$ & $2.04066 \times 10^{35}$ \\
14& 0.0056807& $2.34390\times10^{15}$  & $1.65559\times10^{15}$ & $2.10626\times 10^{35}$\\
16& 0.0055111& $2.39343\times10^{15}$ & $1.68388\times10^{15}$  & $2.17124 \times 10^{35}$\\
18& 0.0053528& $2.44296\times10^{15}$ & $1.71235\times10^{15}$ & $2.23567 \times 10^{35}$\\
20&0.0052048&$2.49250\times10^{15}$ & $1.74099\times10^{15}$ & $2.29960 \times 10^{35}$\\
\hline
\end{tabular*}
\end{table*}

\item{\bf Energy Conditions}\par
For the physical acceptability of the present model, the following energy bounds must be satisfied. NEC (null energy conditions), WEC (weak energy conditions), SEC (strong energy conditions), and DEC (dominant energy conditions) will be satisfied if the following inequalities simultaneously hold:
\begin{eqnarray}
\rho\geq0,\,\rho-p_r\geq 0,\,\rho-p_t\geq 0,\, \rho-p_r-2p_t\geq0.
\end{eqnarray}
The profiles of all these energy conditions are well satisfied as represented graphically in Figures \ref{ec1}-\ref{ec3}. Therefore our present solution is physically viable.

\item{\bf Causality condition and method of ``cracking"}\par
The radial and tangential speeds of sound for our model of compact star are obtained as,
\begin{eqnarray}
  v_r^2 &=& \frac{dp_r}{d\rho}=\beta \\
  v_t^2 &=& \frac{dp_t}{d\rho}=\Big(\frac{dp_t}{dr}\Big)/\Big(\frac{d\rho}{dr}\Big)
\end{eqnarray}
Where the expression for $\frac{d\rho}{dr},\,\frac{dp_r}{dr}$ and $\frac{dp_t}{dr}$ are shown in the expressions (\ref{r1})-(\ref{r3}). For the physically acceptability of the model of relativistic anisotropic star,it should satisfy the causality condition, i.e., the radial and transverse velocity of sound should lie in the range $0<v_r^2,v_t^2<1$. The profile of transverse velocity of sound $(v_t^2)$ is plotted in fig.~\ref{vt} for different values of coupling constant $\alpha$. So from the profile of $v_t^2$, it is clear that the transverse velocity of sound lies in the expected range as well as monotonic decreasing function of r.\par
To check the stability of anisotropic stars under
To check the stability of a fluid sphere, the radial perturbation method was proposed by Herrera \cite{her}, known as ``cracking" method. By using this concept of cracking, Abreu et al. \cite{abreu} proved that the region of an anisotropic fluid sphere where $-1 \leq v_t^2-v_r^2\leq0$ is potentially stable but the region where $0 < v_t^2-v_r^2 \leq 1$ is potentially unstable. For our present model $-1 \leq v_t^2-v_r^2\leq0$ everywhere within the stellar interior and hence our present model is stable.

\item{\bf Relativistic adiabatic index}\par

The stability of a relativistic anisotropic sphere is related to the adiabatic index $\Gamma$ which is the ratio of two specific heats, defined by Chan et al.\cite{chan1},
\begin{equation}
\Gamma_r=\frac{\rho+p_r}{p_r}\frac{dp_r}{d\rho} ~~;~~ \Gamma_t=\frac{\rho+p_t}{p_t}\frac{dp_t}{d\rho}
\end{equation}

Now Bondi \cite{bondi} proposed that $\Gamma>4/3$ gives the condition for the stability of a Newtonian sphere and $\Gamma =4/3$ being the condition for a neutral equilibrium. Now the situation becomes more complicated for an anisotropic general relativistic sphere since the stability depends on the type of anisotropy. We have plotted the profile of $\Gamma_r$ and $\Gamma_t$ in figs. \ref{gammar} and \ref{gammat} respectively for different values of $\alpha$. From these two figures it is clear that both $\Gamma_r,\,\Gamma_t>4/3$ everywhere inside the charged star.

\end{itemize}


\end{document}